\documentclass[aps,reprint,superscriptaddress,nofootinbib,preprintnumbers,longbibliography]{revtex4-1} 

\usepackage{url}
\usepackage{upgreek,amssymb,multirow, mathrsfs}
\usepackage{graphicx,color}
\usepackage{mathtools}
\usepackage{float}
\usepackage{mciteplus}
\usepackage{blindtext}
 
\usepackage{amsmath}
\usepackage{extarrows}
\usepackage{braket}
\usepackage{tikz}
\usepackage{tikz-cd}
\usepackage{bbm}
\usepackage{adjustbox}
\usepackage{xfrac}
\usepackage{appendix}

\usepackage{breakurl}
\usepackage[hidelinks,hyperindex,breaklinks]{hyperref}

\newcommand{\ie}{\begin{equation}\begin{aligned}}
\newcommand{\fe}{\end{aligned}\end{equation}}

\newcommand{\bZ}{{\mathbb{Z}}}

\DeclareRobustCommand
  \cccdots{\mathinner{\cdotp\mkern-3mu\cdotp\mkern-3mu\cdotp}}

\begin{document}

\title{Disentangling anomaly-free symmetries of quantum spin chains}
 
\author{Sahand Seifnashri}
\affiliation{School of Natural Sciences, Institute for Advanced Study, Princeton, NJ} 

\author{Wilbur Shirley}
\affiliation{School of Natural Sciences, Institute for Advanced Study, Princeton, NJ} 
\affiliation{Kadanoff Center for Theoretical Physics, University of Chicago, Chicago, IL}

\begin{abstract}

We clarify the lore that anomaly-free symmetries are either on-site or can be transformed into on-site symmetries. We prove that any finite, internal, anomaly-free symmetry in a 1+1d lattice Hamiltonian system can be disentangled into an on-site symmetry by introducing ancillas and applying conjugation via a finite-depth quantum circuit. We provide an explicit construction of the disentangling circuit using Gauss’s law operators and emphasize the necessity of adding ancillas. Our result establishes the converse to a generalized Lieb-Schultz-Mattis theorem by demonstrating that any anomaly-free symmetry admits a trivially gapped Hamiltonian.

\end{abstract}


\maketitle

\tableofcontents

\section{Introduction}

The notion of \textit{anomaly} is of central importance in quantum field theory and many-body physics. In particular, it serves as a powerful constraint on the dynamics of strongly coupled quantum systems. In lattice systems, such constraints are known as Lieb-Schultz-Mattis (LSM) type constraints \cite{Lieb:1961fr,Affleck:1986pq,Oshikawa:2000zwq,Hastings:2003zx}, while in quantum field theory, they correspond to ’t Hooft anomaly matching conditions \cite{tHooft:1979rat}. (See \cite{McGreevy:2022oyu,Cordova:2022ruw} for reviews.)

The concept of anomaly originates from quantum field theory \cite{Adler:1969gk, Bell:1969ts}, where the \emph{’t Hooft} anomaly of a global symmetry is defined as the obstruction to gauging that symmetry \cite{Kapustin:2014lwa}. The 't Hooft anomaly matching condition states that 't Hooft anomalies match between the high-energy and the effective low-energy theory. In particular, it implies that theories with a non-trivial 't Hooft anomaly cannot be trivially gapped.

In lattice systems, anomalous symmetries typically arise at the boundary of symmetry-protected topological (SPT) phases, via the so-called bulk-boundary correspondence  \cite{Chen:2011pg,Schuch:2011niz,Levin:2012yb,Vishwanath:2012tq,Ryu:2010ah,Wen:2013oza,Kapustin:2014zva}.\footnote{In quantum field theory, this is known as anomaly inflow \cite{Callan:1984sa}.} Specifically, nontrivial SPT states host topologically protected edge modes charged under an anomalous symmetry on the boundary. For instance, the gapless edge modes of 2+1d topological insulators are protected by the combination of charge conservation and time reversal symmetry. From the boundary point of view, the protected edge modes can be understood as a consequence of an \textit{LSM-type} anomaly that can be stated without reference to the bulk.

A symmetry is said to have an LSM-type anomaly if its presence precludes the possibility of a short-range entangled symmetric ground state. In other words, any Hamiltonian that possesses an anomalous symmetry must exhibit one or more of the following features: 1) a gapless spectrum, 2) spontaneous symmetry breaking (SSB), or 3) intrinsic topological order. A classic example is given by the original LSM theorem \cite{Lieb:1961fr}, which states that a translation invariant, SO(3)-symmetric spin-1/2 chain must be gapless or SSB. The LSM theorem is now understood as a consequence of a ``mixed" anomaly between lattice translation and SO(3) symmetry \cite{Cheng:2015kce,Cho:2017fgz,Cheng:2022sgb}, arising from the projective representation of SO(3) per translation unit cell \cite{Chen:2010zpc,ogata2019lieb,Ogata:2020hry}.
In general, it has been shown that anomalies of spin chains pose an obstruction to gauging the symmetry, and thus, are identified as 't Hooft anomalies \cite{Seifnashri:2023dpa}.

Unlike the original LSM anomaly, our focus in this work is on \textit{finite}, \textit{internal} symmetries of quantum spin chains, which may not necessarily possess translation symmetry. We adopt a quantum information-theoretic perspective on anomalies, viewing the anomaly of a symmetry as a form of \textit{entanglement} borne by the symmetry operators. This entanglement is characterized by the anomaly index $[\omega] \in H^3(G, U(1))$, residing in the third group cohomology of the symmetry group $G$ \cite{Chen:2011pg}. A microscopic formula for this index was proposed in \cite{Else:2014vma}. The anomaly index vanishes for \textit{on-site} symmetries and is invariant under the following operations: 
\begin{enumerate}
    \item Tensoring with an on-site $G$-symmetry that acts on an ancillary Hilbert space.
    \item Conjugation by a finite-depth quantum circuit (FDQC).
\end{enumerate}
The cohomology class $[\omega]$ encapsulates the anomaly and, when nontrivial, implies a generalization of the LSM theorem \cite{Kapustin:2024rrm}. The purpose of this Letter is to demonstrate that this index fully characterizes the entanglement structure of an internal symmetry in a 1+1d spin chain.

We accomplish this by explicitly transforming an arbitrary anomaly-free $G$-symmetry into an \textit{on-site} form via the two operations listed above, thereby \textit{disentangling} the symmetry. The latter operation is quite natural and can be thought of as a change of local variables; it may be less clear, however, why the former should be allowed. Physically, ancillas can be thought of as highly inert microscopic/high-energy degrees of freedom lurking throughout the system. In general, it is actually necessary to use this operation to disentangle an anomaly-free symmetry. A simple example is a qubit chain with $\mathbb{Z}_2$-symmetry $U=\prod_j\mathrm{CZ}_{j,j+1}$, where $\mathrm{CZ}_{j,j+1}=(-1)^{(1-Z_j)(1-Z_{j+1})/4}$ is the controlled-$Z$ gate on neighboring qubits. This symmetry is anomaly-free, as it admits the symmetric Hamiltonian $H=-\sum_jZ_j$, but it cannot be disentangled without adding ancillas \cite{Zhang:2024nmp}.

Our result proves the converse to LSM-type theorems for internal finite group symmetries: any anomaly-free symmetry in 1+1d admits a Hamiltonian with a unique short-range entangled gapped ground state (allowing the addition of ancilla degrees of freedom).\footnote{This is the lattice analog of the conjecture in \cite{Seiberg:2019} about continuum quantum field theory, which states that any two theories with the same 't Hooft anomaly must be related by a deformation in which going up and down the renormalization group (RG) flow is allowed. Note that the addition of ancillas here is the analog of going up in the RG flow.} This demonstrates that LSM-type anomalies are equivalent to anomalies as defined by the SPT bulk-boundary correspondence, and labeled by $[\omega]$.

\section{Formal setup}

In this work, we consider spin chains that have a tensor product structure composed of a finite-dimensional local Hilbert space $\mathcal{H}_i$ on each site $i$. On a finite chain $\Lambda$, the total Hilbert space is $\mathcal{H}=\bigotimes_{i\in \Lambda}\mathcal{H}_i$. On infinite chains, it is not sensible to consider such a Hilbert space since it is not separable. Instead, we consider the algebra of local operators constructed from the on-site Hilbert spaces $\mathcal{H}_i$ \cite{Naaijkens:2013vqa}.

The central object in this work is a set of unitary symmetry transformations, $\{U^g\}_{g\in G}$, acting on the local operator algebra of a spin chain. These transformations form a representation of a finite group $G$; that is, they satisfy $U^gU^h=U^{gh}$. We assume these transformations to be locality-preserving in a strict sense, meaning that there is a number $k$ such that, for any operator $\mathcal{O}_i$ acting on a single site $i$, the transformed operator $U^g\mathcal{O}_i(U^g)^\dagger$ acts on sites within the interval $[i-k,i+k]$ for all $g\in G$. The smallest such integer $k$ is called the \emph{range} of the symmetry. By definition, on-site symmetries have range $k=0$, whereas \emph{internal} symmetries have a finite range $k$ that must be independent of $g \in G$.\footnote{For example, the lattice translation operator by $k$ sites is locality-preserving and has range $k$. However, the lattice translation symmetry $\bZ_\mathrm{translation}$ has an infinite range because there is no uniform bound on the entire symmetry group.}
With this definition, any finite, locality-preserving symmetry is necessarily internal.\footnote{A related necessary condition is the triviality of the GNVW index \cite{Gross:2012ygy}. However, the `internal' condition is stronger, as $T \otimes T^{-1}$ has a trivial GNVW index but does not act internally, with $T$ being the lattice translation automorphism.}

On a finite chain, such symmetry transformations are unitary operators acting on the Hilbert space. On an infinite chain, in order to be precise, they must be regarded as automorphisms of the local operator algebra. However, by an abuse of language, we will nonetheless refer to them as symmetry operators. Bearing in mind that automorphisms $U$ and $V$ acting on an infinite number of sites do not have a well-defined overall phase, we will be careful to keep track of the phase ambiguity in equations such as $U=V$. Additionally, we will use `$\propto$' when defining a new local operator that has an inherent phase ambiguity.

Throughout the text, we will focus on the infinite chain $\Lambda = \mathbb{Z}$ to avoid distractions from boundary conditions. However, our results are also valid for finite chains with periodic boundary conditions. We will highlight any subtleties that may arise when extending our results to finite periodic chains.

\section{Background}

Here, we review the formula for the anomaly $[\omega] \in H^3(G,U(1))$ of a symmetry generated by locality-preserving automorphisms $\{U^g\}_{g\in G}$. In particular, we derive identities associated with anomaly-free symmetries, which we will use later for the construction of the disentangler.

\subsection{Sequential circuit and fusion operators}

We begin by constructing a sequential circuit presentation of the symmetry operators $\{U^g\}_{g\in G}$. For an internal symmetry $G$ of range $k$, we consider a choice of truncated symmetry operators $U^g_{\leq I}$ satisfying\footnote{Strictly speaking, $U_{\leq I}^{g}$ is an automorphism of the local operator algebra. A true truncated symmetry operator exists on an interval $[I_0,I]$ for some $I_0 \ll I-2k$.}
\ie
	U_{\leq I}^{g} \, \mathcal{O}_i \, (U_{\leq I}^{g})^\dagger = \begin{cases} U^{g} \, \mathcal{O}_i \, (U^{g})^\dagger & \text{for } i \leq I-k \\
	\, \mathcal{O}_i & \text{for } i > I+k
	\end{cases} \,. \label{truncated.symm}
\fe
Such truncations always exist for internal symmetries acting on spin chains with a finite-dimensional on-site Hilbert space \cite{Seifnashri:2023dpa}.\footnote{See \cite{Gross:2012ygy,Freedman:2019ncn,Farrelly:2019zds} for a general discussion on the obstruction to finding such truncations.} We interpret $U^g_{\leq i}$ as creating a $g$ (topological) defect on link $(i,i+1)$. (See Appendix \ref{sec:defects} for more details on the defect interpretation.)

For simplicity, we introduce a mesoscopic lattice formed by blocks of the original microscopic lattice of size $2k$. We label the sites of the mesoscopic lattice by $j \in \bZ$, which is composed of the block $i\in\{2kj{+}1, 2kj{+}2, \cdots, 2kj{+}2k\} \subset \Lambda$ of the microscopic lattice. Henceforth, we will forget entirely about the original microscopic lattice (whose sites were labeled by $i$) and will only work with the mesoscopic lattice (labeled by $j$). Crucially, the truncated symmetry operators of the mesoscopic lattice respect the condition \eqref{truncated.symm} for $k=\frac12$.

Using the truncations defined above, we obtain the sequential circuit presentation
\ie
    U^g = \cdots U^g_{j+1} U^g_{j} U^g_{j-1} \cdots, \label{sequential}
\fe
where
\ie
    U^g_{j} \propto U^g_{\leq j} (U^g_{\leq j-1})^\dagger \label{movement.op}
\fe
is a \emph{local} unitary operator that implements movement of a $g$ defect from link $(j-1,j)$ to $(j,j+1)$ \cite{Cheng:2022sgb}.\footnote{The right-hand side of \eqref{movement.op} is an automorphism of the operator algebra that determines the unitary operator $U^g_{j}$ up to a phase.}
More generally, we define local unitary \emph{fusion operators} up to an inherent phase ambiguity:
\ie
    \lambda_j(g,h) \propto U^{gh}_{\leq j} \left(U^g_{\leq j-1} U^h_{\leq j}\right)^\dagger. \label{fusion.op}
\fe
This operator implements fusion of a $g$ defect on link $(j{-}1,j)$ and an $h$ defect on link $(j,j{+}1)$ into a $gh$ defect on link $(j,j{+}1)$. Note that $U^g_j=\lambda_j(g,1)$. The fusion operator $\lambda_j(g,h)$ is supported on sites $\{j{-}1,j\}$, therefore $\lambda_j(\cdot\hspace{1pt},\cdot)$ commutes with $\lambda_{j'}(\cdot\hspace{1pt},\cdot)$ for $|j'-j|\ge2$.

It will be convenient to utilize a diagrammatic calculus to represent unitary quantum circuits composed of fusion operators. In our formalism, each diagram has some number of layers, each of which represents a unitary operator. Within a given layer, vertical lines represent the identity operator, and horizontal stacking corresponds to the tensor product of operators. On the other hand, vertical stacking of layers corresponds to the composition of operators. In general, diagrams may be interpreted as spacetime trajectories of symmetry defects. The building blocks of these diagrams are fusion operators and their adjoints:
\ie
\lambda_j(g,h) = \raisebox{-21pt}{
\begin{tikzpicture}[scale=0.7]
  \draw (0.7,0) -- (3.3,0);
  \draw (0.7,1) -- (3.3,1);
  % Draw lattice sites
  \foreach \x in {1,2,3} {
    \filldraw[fill=black, draw=black] (\x,0) circle (0.055);
    \filldraw[fill=black, draw=black] (\x,1) circle (0.055);
  }
  
  % Draw defect
  \draw[color=purple, thick] (1.5,0) -- (1.5,0.5) -- (2.5,0.5);
  \draw[color=purple, thick] (2.5,0) -- (2.5,1);

  % Labels
  \node [below] at (2,0.05) {\scriptsize$j$};
  \node [below] at (1.5,-.1) {\small \color{purple} $g$};
  \node [below] at (2.5,0) {\small \color{purple} $h$};
  \node [above] at (2.5,1) {\small \color{purple} $gh$};
\end{tikzpicture}}\,,~~
\lambda_j^\dagger(g,h) = \raisebox{-21pt}{
\begin{tikzpicture}[scale=0.7]
  \draw (0.7,0) -- (3.3,0);
  \draw (0.7,1) -- (3.3,1);
  % Draw lattice sites
  \foreach \x in {1,2,3} {
    \filldraw[fill=black, draw=black] (\x,0) circle (0.055);
    \filldraw[fill=black, draw=black] (\x,1) circle (0.055);
  }
  
  % Draw defect
  \draw[color=purple, thick] (1.5,1) -- (1.5,0.5) -- (2.5,0.5);
  \draw[color=purple, thick] (2.5,1) -- (2.5,0);

  % Labels
  \node [below] at (2,.05) {\scriptsize$j$};
  \node [above] at (1.5,1) {\small \color{purple} $g$};
  \node [above] at (2.5,1.05) {\small \color{purple} $h$};
  \node [below] at (2.5,0) {\small \color{purple} $gh$};
\end{tikzpicture}}\,.
\fe
As a diagram, the sequential circuit \eqref{sequential} is
\ie U^g = \cdots
\raisebox{-59pt}{
\begin{tikzpicture}[scale=0.7]

\foreach \y in {0,1,2,3,4,5} {
   \draw (0.7,\y) -- (5.3,\y);

  % Draw lattice sites
  \foreach \x in {1,2,3,4,5} {
    \filldraw[fill=black, draw=black] (\x,\y) circle (0.06);
  }}
  
  % Draw defect
  \draw[color=purple, thick] (0.7,0.5) -- (1.5,0.5) -- (1.5,1.5) -- (2.5,1.5) -- (2.5,2.5) -- (3.5,2.5) -- (3.5,3.5) -- (4.5,3.5) -- (4.5,4.5) -- (5.3,4.5);

  % Labels
  \node [below] at (2,0) {\scriptsize$j{-}1$};
  \node [left] at (2.5,2.5) {\small \color{purple} $g$};
  \node [below] at (3,0) {\scriptsize$j$};
  \node [below] at (4,0) {\scriptsize$j{+}1$};
\end{tikzpicture}}\cdots.
\fe

\subsection{The anomaly $[\omega] \in H^3(G, U(1))$}

There exists a simple formula for the anomaly in terms of fusion operators \cite{Seifnashri:2023dpa}, which first appeared in a different form in \cite{Else:2014vma} (see also \cite{kawagoe2021anomalies} for a related method of computation). This formula is expressed in terms of the lattice $F$-symbols, $F_j(g,h,k)$, which are defined diagramatically via an identity referred to as an $F$-move:\footnote{The concept of $F$-symbols in physics originates from the study of RCFTs \cite{Moore:1988qv} and was initially introduced as 6j-symbols in the representation theory of SU(2).}
\ie
\raisebox{-41pt}{
\begin{tikzpicture}[scale=0.7]

\foreach \y in {0,1,2,3} {
  \draw (0.7,\y) -- (4.3,\y);

  % Draw lattice sites
  \foreach \x in {1,2,3,4} {
    \filldraw[fill=black, draw=black] (\x,\y) circle (0.06);
  }}
  
  % Draw defect
  \draw[color=purple, thick] (2.5,0) -- (2.5,0.5) -- (3.5,0.5);
  \draw[color=purple, thick] (1.5,0) -- (1.5,1.5) -- (2.5,1.5) -- (2.5,2.5) -- (3.5,2.5) ;
  \draw[color=purple, thick] (3.5,0) -- (3.5,3);

  % Labels
  \node [below] at (3,.05) {\scriptsize $j$};
  \node [below] at (1.5,-.1) {\small \color{purple} $g$};
  \node [below] at (2.5,0) {\small \color{purple} $h$};
  \node [below] at (3.5,0) {\small \color{purple} $k$};
  \node [above] at (3.5,3) {\small \color{purple} $ghk$};
\end{tikzpicture}} = F_j(g,h,k)
\raisebox{-41pt}{
\begin{tikzpicture}[scale=0.7]

\foreach \y in {0,1,2,3} {
    \draw (0.7,\y) -- (4.3,\y);

  % Draw lattice sites
  \foreach \x in {1,2,3,4} {
    \filldraw[fill=black, draw=black] (\x,\y) circle (0.06);
  }}
  
  % Draw defect
  \draw[color=purple, thick] (1.5,0) -- (1.5,0.5) -- (2.5,0.5);
  \draw[color=purple, thick] (2.5,0) -- (2.5,1.5) -- (3.5,1.5);
  \draw[color=purple, thick] (3.5,0) -- (3.5,3);

  % Labels
  \node [below] at (3,.05) {\scriptsize $j$};
  \node [below] at (1.5,-.1) {\small \color{purple} $g$};
  \node [below] at (2.5,0) {\small \color{purple} $h$};
  \node [below] at (3.5,0) {\small \color{purple} $k$};
  \node [above] at (3.5,3) {\small \color{purple} $ghk$};
\end{tikzpicture}}\,.
\label{equation.for.F}
\fe
As a written equation,
\ie
  &\lambda_{j}(g,hk) \lambda_{j-1}(g,1) \lambda_{j}(h,k) \\=\;&F_j(g,h,k) \lambda_{j}(gh,k)\lambda_{j-1}(g,h).
\fe
It is important to note that $F_j(g,h,k)$ is simply a $U(1)$ phase, since the diagrams on either side of \eqref{equation.for.F} correspond to the same automorphism $U^{ghk}_{\leq j} (U^{g}_{\leq j-2} U^{h}_{\leq j-1} U^{k}_{\leq j})^\dagger$.\footnote{We have used the fact that $\lambda_{j}(h,k)$ is supported on sites $\{j{-}1,j\}$, and thus commutes with the automorphism $(U^{g}_{\leq j-2})^\dagger$.} As shown in \cite{Seifnashri:2023dpa}, the anomaly $[\omega] \in H^3(G,U(1))$ is the cohomology class of the 3-cocycle $\omega_j(g,h,k) = F_j(g,h,k)/F_j(g,h,1)$.\footnote{The idea of computing 't Hooft anomalies using topological defects was emphasized in \cite{Gaiotto:2014kfa}.} Although $\omega_j$ may be $j$-dependent, its cohomology class $[\omega_j] = [\omega]$ is $j$-independent. (See Appendix \ref{app:anomaly} for more details.)

\subsection{Anomaly-free symmetries}

As reviewed in Appendix \ref{app:anomaly}, for anomaly-free symmetries, there is always a choice of phase redefinitions of the fusion operators such that $F_j(g,h,k)=1$ for every site $j$ and $g,h,k\in G$. Henceforth, we will assume such a phase convention. Thus:
\ie
\raisebox{-41pt}{
\begin{tikzpicture}[scale=0.7]

\foreach \y in {0,1,2,3} {
   \draw (0.7,\y) -- (4.3,\y);

  % Draw lattice sites
  \foreach \x in {1,2,3,4} {
    \filldraw[fill=black, draw=black] (\x,\y) circle (0.06);
  }}
  
  % Draw defect
  \draw[color=purple, thick] (2.5,0) -- (2.5,0.5) -- (3.5,0.5);
  \draw[color=purple, thick] (1.5,0) -- (1.5,1.5) -- (2.5,1.5) -- (2.5,2.5) -- (3.5,2.5) ;
  \draw[color=purple, thick] (3.5,0) -- (3.5,3);

  % Labels
  \node [below] at (3,0.05) {\scriptsize $j$};
  \node [below] at (1.5,-.1) {\small \color{purple} $g$};
  \node [below] at (2.5,0) {\small \color{purple} $h$};
  \node [below] at (3.5,0) {\small \color{purple} $k$};
  \node [above] at (3.5,3) {\small \color{purple} $ghk$};
\end{tikzpicture}} =
\raisebox{-41pt}{
\begin{tikzpicture}[scale=0.7]

\foreach \y in {0,1,2,3} {
    \draw (0.7,\y) -- (4.3,\y);

  % Draw lattice sites
  \foreach \x in {1,2,3,4} {
    \filldraw[fill=black, draw=black] (\x,\y) circle (0.06);
  }}
  
  % Draw defect
  \draw[color=purple, thick] (1.5,0) -- (1.5,0.5) -- (2.5,0.5);
  \draw[color=purple, thick] (2.5,0) -- (2.5,1.5) -- (3.5,1.5);
  \draw[color=purple, thick] (3.5,0) -- (3.5,3);

  % Labels
  \node [below] at (3,0.05) {\scriptsize $j$};
  \node [below] at (1.5,-.1) {\small \color{purple} $g$};
  \node [below] at (2.5,0) {\small \color{purple} $h$};
  \node [below] at (3.5,0) {\small \color{purple} $k$};
  \node [above] at (3.5,3) {\small \color{purple} $ghk$};
\end{tikzpicture}}\,. \label{eq:Fmove}
\fe

Using this identity, we obtain an FDQC presentation of the symmetry operators. In particular, as we will show below,
$U^g$ has the following depth-2 circuit presentation:\footnote{For an anomalous symmetry on a finite periodic chain, such an equality holds up to a phase factor determined by the Frobenius-Schur indicator of $g$.}
\ie
    U^g = \prod_{j \text{ odd}} \lambda_j(g,\bar{g}) \prod_{j \text{ even}} \lambda^\dagger_j(\bar{g},g) \label{sequential.fdqc}
\fe
where $\bar{g}=g^{-1}$ is the inverse of $g$. Diagrammatically,
\begin{equation}
    U^g = \cdots
    \raisebox{-31pt}{
    \begin{tikzpicture}[scale=0.7]
    
    \foreach \y in {0,1,2} {
       \draw (0.7,\y) -- (7.3,\y);
    
      % Draw lattice sites
      \foreach \x in {1,2,3,4,5,6,7} {
        \filldraw[fill=black, draw=black] (\x,\y) circle (0.05);
      }}
      
      % Draw defect
      \draw[color=purple, thick] (0.7,1.5) -- (1.5,1.5) -- (1.5,0.5) -- (2.5,0.5) -- (2.5,1.5) -- (3.5,1.5) -- (3.5,0.5)-- (4.5,0.5) -- (4.5,1.5) -- (5.5,1.5) -- (5.5,0.5) -- (6.5,0.5) -- (6.5,1.5) -- (7.5,1.5);
      
      % Labels
      \node [below] at (1,0) {\scriptsize$-3$};
      \node [below] at (2,0) {\scriptsize$-2$};
      \node [below] at (3,0) {\scriptsize$-1$};
      \node [below] at (4,0) {\scriptsize$0$};
      \node [below] at (5,0) {\scriptsize$1$};
      \node [below] at (6,0) {\scriptsize$2$};
      \node [below] at (7,0) {\scriptsize$3$};
      \node [right] at (4.4,1.22) {\small\color{purple} $g$};
      \node [right] at (3.4,1.25) {\small\color{purple} $\bar g$};
      \node [right] at (2.4,1.22) {\small\color{purple} $g$};
      \node [right] at (5.4,1.25) {\small\color{purple} $\bar g$};
      \node [right] at (6.4,1.22) {\small\color{purple} $g$};
      \node [right] at (1.4,1.25) {\small\color{purple} $\bar g$};
    \end{tikzpicture}}
    \cdots.
\end{equation}

\subsection{Gauss's law operators \label{sec:gauss}}

To prove identities involving fusion operators, such as the one in \eqref{sequential.fdqc}, it is useful to define the following set of local operators for each $g \in G$ and site $j$ of the 1D lattice:
\ie
    \mathcal{G}_j^g  = \sum_{a,b,s \in G} \raisebox{-29pt}{ \begin{tikzpicture}[scale=0.7]
\foreach \y in {1,2,3} {
  \draw (0.7,\y) -- (3.3,\y);
  % Draw lattice sites
  \foreach \x in {1,2,3} {
    \filldraw[fill=black, draw=black] (\x,\y) circle (0.06);
  }}
  % Draw defect
  \draw[color=purple, thick] (1.5,1) -- (1.5,1.5) -- (2.5,1.5);
  \draw[color=purple, thick] (2.5,1) -- (2.5,3);
  \draw[color=purple, thick] (1.5,3) -- (1.5,2.5) -- (2.5,2.5);
  % Labels
  \node [below] at (2,1) {\scriptsize $j$};
  \node [below] at (1.5,.9) {\small \color{purple} $a$};
  \node [below] at (2.5,1) {\small \color{purple} $b$};
  \node [above] at (1.5,3) {\small \color{purple} $a\bar{g}$};
  \node [above] at (2.5,3) {\small \color{purple} $gb$};
\end{tikzpicture}} \otimes \ket{s\bar{a},s\bar{g},sb} \bra{s\bar{a},s,sb}_{j-1,j,j+1}.
\label{gauss.operator}
\fe
These operators act on an enlarged Hilbert space $\mathcal{H}\otimes\mathcal{H}'$, where $\mathcal{H}'=\bigotimes_{j\in\mathbb{Z}}\mathbb{C}^{|G|}$ is an ancillary Hilbert space spanned by basis states $\ket{\{s_j\}}=\bigotimes_j \ket{s_j}_j$ with $s_j \in G$. We will refer to $\mathcal{G}_j^g$ as \emph{Gauss's law} operators due to their relation to gauging, which we review in Appendix \ref{app:rel.gauging}.\footnote{We note that, while Gauss's law operators are ordinarily defined in terms of gauge fields residing on lattice links, as in \cite{Seifnashri:2023dpa}, the operators here are defined in terms of ancillary spins living on the \textit{sites} of the lattice.}
There is a natural on-site $G$-symmetry on $\mathcal{H}'$, denoted by $\mathcal{X}^g=\sum_{\{s_j\}}\ket{\{gs_j\}}\bra{\{s_j\}}$.

For an anomaly-free symmetry, the Gauss's law operators satisfy three important properties: they (i) commute at different sites, (ii) form a local $G$-representation, and (iii) commute with $\mathcal{X}^g$. In equations, we have
    \begin{align}
        \mathcal{G}_{j}^{g} \mathcal{G}_{j'}^{h} &= \mathcal{G}_{j'}^{h} \mathcal{G}_{j}^{g} \qquad \text{for} \quad j \neq j', \label{Gauss.commute}\\
        \mathcal{G}_{j}^{g} \mathcal{G}_{j}^{h} &= \mathcal{G}_{j}^{gh},
        \label{Gauss.rep}\\ 
        \mathcal{G}_{j}^{g} \mathcal{X}^h &= \mathcal{X}^h \mathcal{G}_{j}^{g}. \label{Gauss.law}
    \end{align}
The only nontrivial relation above is $\mathcal{G}_{j-1}^g \mathcal{G}_j^h = \mathcal{G}_j^h \mathcal{G}_{j-1}^g$, which follows from \eqref{eq:Fmove} as shown in Appendix \ref{app:gauss.law.commute}.

We now demonstrate \eqref{sequential.fdqc} by rewriting the sequential circuit presentation \eqref{sequential} of $U^g$ in terms of Gauss's law operators:
\ie
    U^g = \bra{\{\bar{g}\}} \cdots \mathcal{G}^{g}_{j+1} \mathcal{G}^{g}_j \mathcal{G}^{g}_{j-1} \cdots \ket{\{1\}}.
\fe
Equation \eqref{sequential.fdqc} is then obtained by taking the $\ket{\{\bar{g}\}}\bra{\{1\}}$ matrix element of the following identity, which itself follows from \eqref{Gauss.commute}:
\ie
    \cdots \mathcal{G}^{g}_{j+1} \mathcal{G}^{g}_j \mathcal{G}^{g}_{j-1} \cdots = \prod_{j \text{ odd}} \mathcal{G}^{g}_j \prod_{j \text{ even}} \mathcal{G}^{g}_j.
\fe

More generally, writing $\prod_j \mathcal{G}_{j}^{g_j}$ in different orders of multiplication produces infinitely many identities involving circuits made of fusion operators.\footnote{In particular, two diagrams represent the same quantum circuit if they correspond to flat spacetime gauge field configurations with the same holonomies, i.e., if they are related by a spacetime gauge transformation.}

\section{The disentangler}

In this section, we construct an explicit \textit{disentangler} for an arbitrary anomaly-free $G$-symmetry $\{U^g\}_{g\in G}$, using fusion operators as building blocks.
Specifically, given a choice of fusion operators satisfying \eqref{eq:Fmove}, we construct a finite-depth circuit $\mathcal{W}$, acting on the enlarged Hilbert space $\mathcal{H}\otimes\mathcal{H}'$ described above, that satisfies
\begin{equation}
    \mathcal{W}^\dagger(U^g\otimes\mathcal{X}^g)\mathcal{W}=\mathbbm{1} \otimes\mathcal{X}^g,
\end{equation}
where $\mathcal{X}^g$ is an on-site $G$-symmetry acting on the ancillary Hilbert space $\mathcal{H}'$. To motivate the general solution, we begin with the simple case $G=\mathbb{Z}_2$.

\subsection{The $G=\mathbb{Z}_2$ case}
Denote the elements of $\mathbb{Z}_2$ by $\{0,1\}$. We consider an anomaly-free $\mathbb{Z}_2$ symmetry generated by a unitary $U$ obeying the multiplication law $U^2=1$.

As a warm-up, we first consider a baby version of the problem: ignoring locality, can $U$ be transformed into a single-qubit Pauli operator? There is a simple solution. First, add an ancilla qubit on an arbitrary site, enlarging the Hilbert space from $\mathcal{H}$ to $\mathcal{H}\otimes\mathbb{C}^2$. Moreover, extend the symmetry from $U$ to $U\otimes X$. Then, define a unitary transformation
\ie
    W=\mathbbm{1}\otimes\ket{0}\bra{0}+U\otimes\ket{1}\bra{1},
\fe
which, in the quantum computation parlance, constitutes a controlled-$U$ gate. $W$ maps $U\otimes X$ to the single-qubit Pauli operator $\mathbbm{1}\otimes X$:
\ie
    W (U \otimes X ) W^\dagger = \mathbbm{1} \otimes X\,,
\fe
thereby accomplishing the task.

The transformation $W$ violates locality because it couples the ancilla qubit to every site of the 1D lattice. To restore locality, we instead distribute ancillas throughout the 1+1d system; in particular, we add one qubit to each site $j$ of the coarse-grained lattice, spanned by basis states $\ket{0}_j$ and $\ket{1}_j$. The Hilbert space is thereby enlarged from $\mathcal{H}$ to $\mathcal{H}\otimes\mathcal{H}'$ where $\mathcal{H}'=\bigotimes_j\mathbb{C}^2$. Moreover, we extend $U$ to $U\otimes\mathcal{X}$ where $\mathcal{X}=\prod_jX_j$. This allows us to upgrade $W$ to a sequential circuit
\ie
    \mathcal{W} = \cdots \mathcal{W}_{j+1}\mathcal{W}_j\mathcal{W}_{j-1} \cdots,
    \label{eq:Z2W}
\fe
where
\ie
    \mathcal{W}_j =\mathbbm{1}\otimes\ket{0}\bra{0}_j&+\lambda_j(1,0)\otimes\ket{11}\bra{11}_{j,j+1}\\&+\lambda_j(1,1)\otimes\ket{10}\bra{10}_{j,j+1},
\fe
in terms of the the nontrivial $G=\mathbb{Z}_2$ fusion operators:
\ie
\lambda_j(1,0)=\raisebox{-17pt}{
\begin{tikzpicture}[scale=0.5]
  \draw (0.7,0) -- (3.3,0);
  \draw (0.7,1) -- (3.3,1);
  % Draw lattice sites
  \foreach \x in {1,2,3} {
    \filldraw[fill=black, draw=black] (\x,0) circle (0.055);
    \filldraw[fill=black, draw=black] (\x,1) circle (0.055);
  }  
  % Draw defect
  \draw[color=purple, thick] (1.5,0) -- (1.5,0.5) -- (2.5,0.5) -- (2.5,1);
  % Labels
  \node [below] at (2,0.05) {\scriptsize $j$};
\end{tikzpicture}},~~\lambda_j(1,1)=
\raisebox{-17pt}{
\begin{tikzpicture}[scale=0.5]
  \draw (0.7,0) -- (3.3,0);
  \draw (0.7,1) -- (3.3,1);
  % Draw lattice sites
  \foreach \x in {1,2,3} {
    \filldraw[fill=black, draw=black] (\x,0) circle (0.055);
    \filldraw[fill=black, draw=black] (\x,1) circle (0.055);
  }  
  % Draw defect
  \draw[color=purple, thick] (1.5,0) -- (1.5,0.5) -- (2.5,0.5) -- (2.5,0);
  % Labels
  \node [below] at (2,0.05) {\scriptsize $j$};
\end{tikzpicture}}.
\fe

In these diagrams, red lines correspond to defects of the nonzero element of $\mathbb{Z}_2$. Although $\mathcal{W}$ is expressed as a sequential circuit, it is actually an FDQC, as will be demonstrated through the general arguments given in the next section. $\mathcal{W}$ is a disentangler for the $\mathbb{Z}_2$ symmetry, since
\ie
    \mathcal{W}^\dagger(U\otimes\mathcal{X})\mathcal{W}=\mathbbm{1} \otimes\mathcal{X} \,.
    \label{eq:UX}
\fe

To derive this transformation law, express $\mathcal{W}$ in \eqref{eq:Z2W} in the following manner:
\begin{equation}
    \mathcal{W}=\sum_{\{s_j\}}\Lambda_{\{s_j\}}\otimes\ket{\{s_j\}}\bra{\{s_j\}} \,,
\end{equation}
where $s_j \in\mathbb{Z}_2$ is the state of the ancilla qubit on site $j$, and $\Lambda_{\{s_j\}}$ is the following sequential circuit:\footnote{This sequential circuit has ordering $\cdots \lambda_{j+1} \lambda_j \lambda_{j-1} \cdots$.}
\ie
    \Lambda_{\{s_j\}} =\prod_j\lambda_j(s_j,s_j+s_{j+1}).
\fe
To illustrate the meaning of $\Lambda_{\{s_j\}}$, consider a sample qubit configuration in which $s_1=s_2=1$ and $s_j=0$ otherwise. In this case,
\ie
\Lambda_{\{s_j\}}=
\raisebox{-23pt}{
\begin{tikzpicture}[scale=0.5]

\foreach \y in {0,1,2} {
   \draw (1.7,\y) -- (5.3,\y);

  % Draw lattice sites
  \foreach \x in {2,3,4,5} {
    \filldraw[fill=black, draw=black] (\x,\y) circle (0.06);
  }}
  
  % Draw defect
  \draw[color=purple, thick] (2.5,0) -- (2.5,0.5) -- (3.5,0.5) -- (3.5,1.5) -- (4.5,1.5) -- (4.5,0);

  % Labels
  \node [below] at (3,0) {\scriptsize$1$};
  \node [below] at (4,0) {\scriptsize$2$};
\end{tikzpicture}}\,,~~
\Lambda_{\{1+s_j\}}=\cccdots
\raisebox{-51pt}{
\begin{tikzpicture}[scale=0.5]

\foreach \y in {0,1,2,3,4,5,6} {
   \draw (0.7,\y) -- (6.3,\y);

  % Draw lattice sites
  \foreach \x in {1,2,3,4,5,6} {
    \filldraw[fill=black, draw=black] (\x,\y) circle (0.06);
  }}
  
  % Draw defect
  \draw[color=purple, thick] (4.5,0) -- (4.5,4.5) -- (5.5,4.5) -- (5.5,5.5) -- (6.3,5.5);
  \draw[color=purple, thick] (0.7,0.5) -- (1.5,0.5) -- (1.5,1.5) -- (2.5,1.5) -- (2.5,0);

  % Labels
  \node [below] at (3,0) {\scriptsize$1$};
  \node [below] at (4,0) {\scriptsize$2$};
\end{tikzpicture}}\cccdots.
\fe

The transformation law (\ref{eq:UX}) then follows from the relation
\ie
    U\Lambda_{\{s_j\}}=\Lambda_{\{1+s_j\}} \,,
    \label{Lambda.relation.Z2}
\fe
which in turn is a consequence of the two nontrivial $F$-moves obtained by substituting $g=h=1$ in \eqref{eq:Fmove}:
\ie
\raisebox{-32pt}{
\begin{tikzpicture}[scale=0.5]

\foreach \y in {0,1,2,3} {
   \draw (0.7,\y) -- (4.3,\y);

  % Draw lattice sites
  \foreach \x in {1,2,3,4} {
    \filldraw[fill=black, draw=black] (\x,\y) circle (0.06);
  }}
  
  % Draw defect
  \draw[color=purple, thick] (2.5,0) -- (2.5,0.5) -- (3.5,0.5);
  \draw[color=purple, thick] (1.5,0) -- (1.5,1.5) -- (2.5,1.5) -- (2.5,2.5) -- (3.5,2.5) ;
  \draw[color=purple, thick] (3.5,0.5) -- (3.5,2.5);

  % Labels
  \node [below] at (3,0) {\scriptsize$j$};
\end{tikzpicture}}\,\,=
\raisebox{-17.75pt}{
\begin{tikzpicture}[scale=0.5]

\foreach \y in {0,1} {
    \draw (0.7,\y) -- (3.3,\y);

  % Draw lattice sites
  \foreach \x in {1,2,3} {
    \filldraw[fill=black, draw=black] (\x,\y) circle (0.06);
  }}
  
  % Draw defect
  \draw[color=purple, thick] (1.5,0) -- (1.5,0.5) -- (2.5,0.5);
  \draw[color=purple, thick] (2.5,0) -- (2.5,0.5);

  % Labels
  \node [below] at (3,0) {\scriptsize$j$};
\end{tikzpicture}}\,,~~
\raisebox{-32pt}{
\begin{tikzpicture}[scale=0.5]

\foreach \y in {0,1,2,3} {
   \draw (0.7,\y) -- (4.3,\y);

  % Draw lattice sites
  \foreach \x in {1,2,3,4} {
    \filldraw[fill=black, draw=black] (\x,\y) circle (0.06);
  }}
  
  % Draw defect
  \draw[color=purple, thick] (2.5,0) -- (2.5,0.5) -- (3.5,0.5);
  \draw[color=purple, thick] (1.5,0) -- (1.5,1.5) -- (2.5,1.5) -- (2.5,2.5) -- (3.5,2.5) ;
  \draw[color=purple, thick] (3.5,0) -- (3.5,0.5);
  \draw[color=purple, thick] (3.5,2.5) -- (3.5,3);

  % Labels
  \node [below] at (3,0) {\scriptsize$j$};
\end{tikzpicture}}\,\,=
\raisebox{-17.75pt}{
\begin{tikzpicture}[scale=0.5]

\foreach \y in {0,1} {
    \draw (0.7,\y) -- (4.3,\y);

  % Draw lattice sites
  \foreach \x in {1,2,3,4} {
    \filldraw[fill=black, draw=black] (\x,\y) circle (0.06);
  }}
  
  % Draw defect
  \draw[color=purple, thick] (1.5,0) -- (1.5,0.5) -- (2.5,0.5);
  \draw[color=purple, thick] (2.5,0) -- (2.5,0.5);
  \draw[color=purple, thick] (3.5,0) -- (3.5,1);

  % Labels
  \node [below] at (3,0) {\scriptsize$j$};
\end{tikzpicture}}\,.\label{eq:FmoveZ2}
\fe
For instance, suppose again that $s_1=s_2=1$ and $s_j=0$ otherwise. In this case,\footnote{Here, we have used the fact that $\lambda_j(\cdot\hspace{1pt},\cdot)$ commutes with $\lambda_{j'}(\cdot\hspace{1pt},\cdot)$ for $|j-j'|\ge2$ to shift some of the gates vertically as long as they do not touch other parts of the diagram.}
\begin{equation}
\begin{alignedat}{1}
&U\Lambda_{\{s_j\}}=\\[5pt]&\cccdots
\raisebox{-44pt}{
\begin{tikzpicture}[scale=0.5]

\foreach \y in {0,1,2,3,4,5} {
   \draw (1.7,\y) -- (5.3,\y);

  % Draw lattice sites
  \foreach \x in {2,3,4,5} {
    \filldraw[fill=black, draw=black] (\x,\y) circle (0.06);
  }}
  
  % Draw defect
  \draw[color=purple, thick] (2.5,0) -- (2.5,0.5) -- (3.5,0.5) -- (3.5,1.5) -- (4.5,1.5) -- (4.5,0);
  \draw[color=purple, thick] (1.7,1.5) -- (2.5,1.5) -- (2.5,2.5) -- (3.5,2.5) -- (3.5,3.5) -- (4.5,3.5) -- (4.5,4.5) -- (5.3,4.5);

  % Labels
  \node [below] at (3,0) {\scriptsize$1$};
  \node [below] at (4,0) {\scriptsize$2$};
\end{tikzpicture}}\cccdots=\cccdots
\raisebox{-44pt}{
\begin{tikzpicture}[scale=0.5]

\foreach \y in {0,1,2,3,4,5} {
   \draw (1.7,\y) -- (5.3,\y);

  % Draw lattice sites
  \foreach \x in {2,3,4,5} {
    \filldraw[fill=black, draw=black] (\x,\y) circle (0.06);
  }}
  
  % Draw defect
  \draw[color=purple, thick] (4.5,0) -- (4.5,4.5) -- (5.3,4.5);
  \draw[color=purple, thick] (1.7,1.5) -- (2.5,1.5) -- (2.5,2.5) -- (3.5,2.5)-- (3.5,0.5) -- (2.5,0.5) -- (2.5,0);

  % Labels
  \node [below] at (3,0) {\scriptsize$1$};
  \node [below] at (4,0) {\scriptsize$2$};
\end{tikzpicture}}\cccdots=\cccdots
\raisebox{-44pt}{
\begin{tikzpicture}[scale=0.5]

\foreach \y in {0,1,2,3,4,5} {
   \draw (1.7,\y) -- (5.3,\y);

  % Draw lattice sites
  \foreach \x in {2,3,4,5} {
    \filldraw[fill=black, draw=black] (\x,\y) circle (0.06);
  }}
  
  % Draw defect
  \draw[color=purple, thick] (4.5,0) -- (4.5,4.5) -- (5.3,4.5);
  \draw[color=purple, thick] (1.7,1.5) -- (2.5,1.5) -- (2.5,0);

  % Labels
  \node [below] at (3,0) {\scriptsize$1$};
  \node [below] at (4,0) {\scriptsize$2$};
\end{tikzpicture}}\cccdots
\\&\hspace{197pt}=\Lambda_{\{1+s_j\}}.
\end{alignedat}
\end{equation}
Similar reasoning can be used to demonstrate \eqref{Lambda.relation.Z2} for other $\{s_j\}$ configurations. For a general argument, see the following section.

In certain special cases, the $U_j$ operators may form commuting representations of $\mathbb{Z}_2$, that is, $(U_j)^2=1$ and $[U_{j},U_{j+1}]=0$. As a result, $\lambda_j(1,1)=\lambda_j(1,0)=U_j$, so $\mathcal{W}$ simplifies to a product of controlled-$U_j$ operators:
\ie
    \mathcal{W}=&\prod_j\Big(\mathbbm{1}\otimes\ket{0}\bra{0}_j+U_j\otimes\ket{1}\bra{1}_j\Big).
\fe
For example, consider a qubit chain with the $\mathrm{CZ}$ symmetry $U=\prod_j\mathrm{CZ}_{j,j+1}$. Introduce an ancilla qubit on each site $j$, whose Pauli operators are denoted by $Z_j',X_j'$. Setting $U_j=\mathrm{CZ}_{j,j+1}$, it follows that $\mathcal{W}=\prod_j\mathrm{CCZ}_j$, where $\mathrm{CCZ}_j=(-1)^{(1-Z_j')(1-Z_j)(1-Z_{j+1})/8}$ is the controlled-$\mathrm{CZ}$ gate on the ancilla on site $j$ and physical qubits on sites $j,j+1$. Since $\mathrm{CCZ}_j$ maps $\mathrm{CZ}_{j,j+1}\otimes X_j'\to \mathbbm{1}\otimes X_j'$, we recover the transformation law (\ref{eq:UX}).

\subsection{Finite group $G$ \label{sec:disentangling.G}}

We now generalize this solution to construct a disentangler for an arbitrary anomaly-free $G$-symmetry, for any finite group $G$. Instead of an ancilla qubit, we add a $|G|$-dimensional ancilla qudit to each site $j$ of the 1D lattice. Thus, the Hilbert space is enlarged from $\mathcal{H}$ to $\mathcal{H}\otimes\mathcal{H}'$, where $\mathcal{H}'=\bigotimes_{j\in\mathbb{Z}}\mathbb{C}^{|G|}$ as in Section \ref{sec:gauss}. The ancillary Hilbert space $\mathcal{H}'$ is spanned by basis states denoted $\ket{\{s_j\}}=\bigotimes_j \ket{s_j}_j$ with $s_j \in G$.
There is a natural on-site $G$-symmetry action on $\mathcal{H}'$, generated by $\mathcal{X}^g=\prod_jL^g_j=\sum_{\{s_j\}}{\ket{\{gs_j\}}}{\bra{\{s_j\}}}$ where $L_j^g = \sum_s {\ket{gs}}{\bra{s}}_j$ is the left multiplication operator acting on the qudit at site $j$.

The disentangler is written in the form of a sequential circuit, generalizing the expression in (\ref{eq:Z2W}):
\ie
    \mathcal{W} = \cdots \mathcal{W}_{j+1}\mathcal{W}_j\mathcal{W}_{j-1} \cdots,
\fe
where
\ie
    \mathcal{W}_j = \sum_{s,t\in G}\lambda_j(s,\bar{s}t)\otimes\ket{s,t}\bra{s,t}_{j,j+1} \,.
\fe
This operator is engineered to satisfy
\begin{equation}
    \mathcal{W}^\dagger(U^g\otimes\mathcal{X}^g)\mathcal{W}=\mathbbm{1} \otimes\mathcal{X}^g \, .
    \label{eq:UgXg}
\end{equation}

To derive this transformation law, we again express $\mathcal{W}$ as follows:
\begin{equation}
    \mathcal{W}=\sum_{\{s_j\}}\Lambda_{\{s_j\}}\otimes\ket{\{s_j\}}\bra{\{s_j\}} \,,
\end{equation}
where, using the Gauss's law operators of Section \ref{sec:gauss},
\ie
    \Lambda_{\{s_j\}} &=  \big \langle \{1\} \big| \cdots \mathcal{G}^{s_{j+1}}_{j+1} \mathcal{G}^{s_{j}}_j \mathcal{G}^{s_{j-1}}_{j-1} \cdots \big|\{s_j\} \big\rangle \\
    &=\prod_j\lambda_j(s_j,\bar{s}_js_{j+1})\\
    &= \cdots
    \raisebox{-40pt}{
    \begin{tikzpicture}[scale=0.7]
    
    \foreach \y in {0,1,2,3} {
       \draw (0.7,\y) -- (5.3,\y);
    
      % Draw lattice sites
      \foreach \x in {1,2,3,4,5} {
        \filldraw[fill=black, draw=black] (\x,\y) circle (0.06);
      }}
      
      % Draw defect
      \draw[color=purple, thick] (1.5,0) -- (1.5,0.5) -- (2.5,0.5);
      \draw[color=purple, thick] (2.5,0) -- (2.5,1.5) -- (3.5,1.5);
      \draw[color=purple, thick] (3.5,0) -- (3.5,2.5) -- (4.5,2.5);
      \draw[color=purple, thick] (4.5,0) -- (4.5,3);
      
      % Labels
      \node [above] at (3,3) {\scriptsize$1$};
      \node [below] at (1.5,-0.05) {\small\color{purple} $s_0$};
      \node [below] at (2.5,0) {\small\color{purple} $\bar{s}_0s_1$};
      \node [below] at (3.5,0) {\small\color{purple} $\bar{s}_1s_2$};
      \node [below] at (4.5,0) {\small\color{purple} $\bar{s}_2s_3$};
      \node [above] at (4.5,3) {\small\color{purple} $s_3$};
    \end{tikzpicture}} \cdots \,.
    \label{Lambda}
\fe
Then, (\ref{eq:UgXg}) follows from the relation
\ie
    U^g\Lambda_{\{s_j\}}=\Lambda_{\{gs_j\}},
    \label{Lambda.relation}
\fe
which we prove in two different ways. For one, we can obtain this relation via an infinite sequence of $F$-moves. For instance, one move of this sequence is depicted below:
\ie
\raisebox{-51pt}{
\begin{tikzpicture}[scale=0.7]

\foreach \y in {0,1,2,3,4} {
   \draw (0.7,\y) -- (5.3,\y);

  % Draw lattice sites
  \foreach \x in {1,2,3,4,5} {
    \filldraw[fill=black, draw=black] (\x,\y) circle (0.06);
  }}
  
  % Draw defect
  \draw[color=purple, thick] (2.5,0) -- (2.5,0.5) -- (3.5,0.5);
  \draw[color=purple, thick] (3.5,0) -- (3.5,1.5) -- (4.5,1.5);
  \draw[color=purple, thick] (4.5,0) -- (4.5,4);
  \draw[color=teal, thick] (1.5,0) --(1.5,1.5) -- (2.5,1.5) -- (2.5,2.5) -- (3.5,2.5) -- (3.5,3.5) -- (4.5,3.5);

  % Labels
  \node [below] at (1.5,-.05) {\small\color{teal}$g$};
  \node [above] at (3,4) {\scriptsize$1$};
  \node [below] at (2.5,-0.05) {\small\color{purple} $s_1$};
  \node [below] at (3.5,0) {\small\color{purple} $\bar{s}_1s_2$};
  \node [below] at (4.6,0) {\small\color{purple} $\bar{s}_2s_3$};
  \node [above] at (4.5,4) {\small\color{purple} $gs_3$};
\end{tikzpicture}} \,=
\raisebox{-51pt}{
\begin{tikzpicture}[scale=0.7]

\foreach \y in {0,1,2,3,4} {
   \draw (0.7,\y) -- (5.3,\y);

  % Draw lattice sites
  \foreach \x in {1,2,3,4,5} {
    \filldraw[fill=black, draw=black] (\x,\y) circle (0.06);
  }}
  
  % Draw defect
  \draw[color=purple, thick] (2.5,0) -- (2.5,0.5) -- (3.5,0.5);
  \draw[color=purple, thick] (3.5,0) -- (3.5,3.5) -- (4.5,3.5);
  \draw[color=purple, thick] (4.5,0) -- (4.5,4);
  \draw[color=teal, thick] (1.5,0) --(1.5,1.5) -- (2.5,1.5) -- (2.5,2.5) -- (3.5,2.5);

  % Labels
  \node [below] at (1.5,-.05) {\small\color{teal}$g$};
  \node [above] at (3,4) {\scriptsize$1$};
  \node [below] at (2.4,-0.05) {\small\color{purple} $s_1$};
  \node [below] at (3.5,0) {\small\color{purple} $\bar{s}_1s_2$};
  \node [below] at (4.6,0) {\small\color{purple} $\bar{s}_2s_3$};
  \node [above] at (4.5,4) {\small\color{purple} $gs_3$};
\end{tikzpicture}} \,.
\fe

Alternatively, we can prove \eqref{Lambda.relation} by leveraging the properties of the Gauss's law operators in (\ref{Gauss.commute}-\ref{Gauss.law}). We first express $\Lambda_{\{s_j\}}$ and $U^g$ in terms of Gauss's law operators as follows:
\ie
    U^g&= \bra{\{\bar{g}\}} \prod_j \mathcal{G}_{j}^{g} \ket{\{1\}}\,, \\
    \Lambda_{\{s_j\}} &= \bra{\{1\}}\prod_j \mathcal{G}_{j}^{s_j} \ket{\{s_j\}} \,.
    \label{disentangler.gauss}
\fe
Then, applying \eqref{Gauss.commute} and \eqref{Gauss.rep}, we find that
\ie
    U^g \Lambda_{\{s_j\}} &= \bra{\{\bar{g}\}} \prod_j \mathcal{G}_{j}^{g s_j} \ket{\{s_j\}} \\
    &= \bra{\{1\}} \prod_j \mathcal{G}_{j}^{g s_j} \ket{\{gs_j\}} = \Lambda_{\{gs_j\}} \,.
\fe
The second equality is obtained using \eqref{Gauss.law}.

To complete the discussion, it remains to show that $\mathcal{W}$ is actually an FDQC. To do so, we use Gauss's law operators to express $\Lambda_{\{s_j\}}$ as a depth-2 circuit acting on $\mathcal{H}$:\footnote{It is also possible to obtain the final expression via an infinite sequence of $F$-moves starting from \eqref{Lambda}.}
\ie
    \Lambda_{\{s_j\}} &=  \big \langle \{1\} \big| \prod_{j \text{ even}} \mathcal{G}_j^{s_j} \prod_{j \text{ odd}} \mathcal{G}_j^{s_j} \big|\{s_j\} \big\rangle \\
    &= \cdots
    \raisebox{-40pt}{
    \begin{tikzpicture}[scale=0.7]
    
    \foreach \y in {0,1,2,3} {
       \draw (0.7,\y) -- (7.3,\y);
    
      % Draw lattice sites
      \foreach \x in {1,2,3,4,5,6,7} {
        \filldraw[fill=black, draw=black] (\x,\y) circle (0.06);
      }}
      
      % Draw defect
      \draw[color=purple, thick] (0.7,2.5) -- (1.5,2.5) -- (1.5,1.5) -- (2.5,1.5);
      \draw[color=purple, thick] (1.5,0) -- (1.5,0.5) -- (2.5,0.5);
      \draw[color=purple, thick] (2.5,0) -- (2.5,2.5) -- (3.5,2.5) -- (3.5,1.5) -- (4.5,1.5);
      \draw[color=purple, thick] (3.5,0) -- (3.5,0.5) -- (4.5,0.5);
      \draw[color=purple, thick] (4.5,0) -- (4.5,2.5) -- (5.5,2.5) -- (5.5,1.5) -- (6.5,1.5);
      \draw[color=purple, thick] (5.5,0) -- (5.5,0.5) -- (6.5,0.5);
      \draw[color=purple, thick] (6.5,0) -- (6.5,2.5) -- (7.3,2.5);
      
      % Labels
      \node [above] at (4,3) {\scriptsize$1$};
      \node [below] at (2.5,-0.1) {\small\color{purple} $\cdots$};
      \node [below] at (3.5,0) {\small\color{purple} $\bar{s}_0s_1$};
      \node [below] at (4.5,0) {\small\color{purple} $\bar{s}_1s_2$};
      \node [below] at (5.5,-0.1) {\small\color{purple} $\cdots$};
    \end{tikzpicture}}
    \cdots \,.
\fe
This allows us to express $\mathcal{W}$ as the following FDQC:\footnote{In this expression, each factor is regarded as a single gate. The gates in the product over even $j$ mutually commute; however, neighboring gates act on overlapping ancillas, so the product must be split into two layers. Thus, this expression constitutes a depth-3 circuit on $\mathcal{H}\otimes\mathcal{H}'$.}
\begin{equation}
\begin{split}
    &\mathcal{W}=\prod_{j\text{ even}}\sum_{s\in G}\lambda_j(s,\bar{s})\otimes\ket{s}\bra{s}_j\\&\times\prod_{j\text{ odd}}\sum_{r,s,t\in G}\lambda_j^\dagger(\bar{r},t)\lambda_j(\bar{r}s,\bar{s}t)\otimes\ket{r,s,t}\bra{r,s,t}_{j-1,j,j+1}.
\end{split}
\end{equation}

\section{Conclusion}

In this work, we have studied finite-group internal symmetries of spin chains. Given a symmetry with a trivial anomaly index $[\omega]\in H^3(G,U(1))$, we have constructed an explicit disentangler that transforms each symmetry operator into an on-site form. We note that this construction can also be used to transform between different anomalous symmetries with the same anomaly index, due to the additivity of the index under stacking. Thus, our results may be regarded as a theory of finite group representations on quantum spin chains.\footnote{This theory may further be regarded as a generalization of the theory of projective group representations, which are classified by an index in $H^2(G,U(1))$ and physically correspond to anomalous symmetries on the 0+1d boundary of a 1+1d system (which is composed of two well-separated points).}

Our result provides a recipe for constructing a 2+1d bulk SPT that realizes a given anomalous spin chain $G$-symmetry as the effective symmetry of its low-energy edge theory. In doing so, it fills a gap in the literature by furnishing the boundary-to-bulk direction of the bulk-boundary correspondence. The recipe is as follows: given a $G$-symmetry $\{U^g\}_{g\in G}$ on a 1D spin chain with anomaly index $[\omega] \in H^3(G,U(1))$, we construct an SPT entangler on a 2D system formed by an infinite array of 1D chains, each labeled by an integer $k$, tensored with ancillas transforming under an on-site $G$-symmetry distributed throughout each chain. The 1D chains are arranged with alternating orientation, positive for even $k$ and negative for odd $k$, such that the anomaly index on odd chains is $\omega^{-1}$ rather than $\omega$. Denote the on-site symmetry operator on chain $k$ by $\mathcal{X}^g_k$, and let $V^g_k=U^g_k\otimes\mathcal{X}^g_k$. The SPT entangler we construct creates an SPT state protected by the total on-site symmetry $\mathcal{X}^g=\prod_k\mathcal{X}^g_k$. It is composed of two layers: the first layer is composed of an array of finite-depth quantum circuits mapping $\mathcal{X}^g_{2k-1}\otimes\mathcal{X}^g_{2k}\to V^g_{2k-1}\otimes V^g_{2k}$, and the second layer is composed of an array of finite-depth quantum circuits mapping $V^g_{2k}\otimes V^g_{2k+1}\to \mathcal{X}^g_{2k}\otimes\mathcal{X}^g_{2k+1}$. Clearly, this two-layer circuit preserves the symmetries $\mathcal{X}^g$. In a finite system, it `pumps' an anomalous $G$-symmetry from one edge to the other, hence it is an SPT entangler.

It is natural to ask how our results may be generalized. We expect that they can be extended to continuous symmetries of spin chains. In higher dimensions, there are additional indices beyond the anomaly index that prevent arbitrary anomaly-free symmetries from being disentangled \cite{H2}. Understanding the total obstruction to disentangle-ability is an interesting challenge for future work. Furthermore, it would be interesting to explore the notions of on-site symmetry and disentangling in the context of generalized symmetries \cite{Gaiotto:2014kfa,McGreevy:2022oyu,Cordova:2022ruw}.

\begin{acknowledgments}

We are grateful to Xie Chen, Lukasz Fidkwoski, Michael Hermele, Ruochen Ma, Nikita Sopenko, Ryan Thorngren, and, especially, Michael Levin for interesting discussions. SS gratefully acknowledges support from the Sivian Fund and the Paul Dirac Fund at the Institute for Advanced Study, the U.S. Department of Energy grant DE-SC0009988, and the Simons Collaboration on Ultra-Quantum Matter, which is a grant from the Simons Foundation (651444, NS). WS gratefully acknowledges support from the Kadanoff Center for Theoretical Physics. The authors of this paper were ordered alphabetically.

\end{acknowledgments}

\onecolumngrid
\appendix

\section{More on the anomaly $[\omega] \in H^3(G,U(1))$ \label{app:anomaly}}

In this Appendix, we review the precise relation between the lattice $F$-symbol $F_j(g,h,k)$, defined in \eqref{equation.for.F}, and the anomaly index $[\omega] \in H^3(G,U(1))$. In particular, we show that when the symmetry is anomaly-free, the phases $F_j(g,h,k)$ can be made trivial by appropriate phase redefinition of the fusion operators $\lambda_j(g,h)$.

We begin by establishing the terminology and notation for group cohomology. Let $C^n(G,U(1))$ denote the set of all $n$-cochains, that is, all functions $f : G^n \to U(1)$. These cochains form an abelian group under pointwise multiplication, i.e., $(f\cdot e)(g_1,g_2,\dots,g_n)=f(g_1,g_2,\dots,g_n) e(g_1,g_2,\dots,g_n)$. The coboundary map on the set of cochains is defined by
\ie
    {}& \delta_n : C^n(G,U(1)) \to C^{n+1}(G,U(1)) \,, \\
    & \left( \delta_n f \right)(g_1,g_2, \dots , g_{n+1}) = f(g_2,\dots,g_{n+1}) f^{(-1)^{n+1}}(g_1,\dots,g_n) \prod_{i=1}^n f^{(-1)^i}(g_1,\dots,g_{i-1},g_i g_{i+1}, \dots,g_{n+1}) \,,
\fe
where $f^{(-1)} = 1/f$. The $n$-th cohomology group of $G$ is given by
\ie
    H^n(G,U(1)) = \mathrm{kernel}(\delta_n) / \mathrm{image}(\delta_{n-1}) \,.
\fe

From the lattice $F$-symbols, we define the 3-cochain $\omega_j \in C^3(G, U(1))$ and 2-cochain $\alpha_j \in C^2(G, U(1))$:
\ie
    \omega_j(g_1,g_2,g_3) = \frac{F_j(g_1,g_2,g_3)}{F_j(g_1,g_2,1)} \,, \qquad  \alpha_j(g_1,g_2) = F_j(g_1,g_2,1) \,.
\fe
As shown in \cite{Seifnashri:2023dpa}, the lattice $F$-symobls satisfy a modified pentagon/cocycle equation:
\ie
    \frac{F_j(g_2,g_3,g_4)F_j(g_1,g_2g_3,g_4)F_{j-1}(g_1,g_2,g_3)}{F_j(g_1g_2,g_3,g_4)F_j(g_1,g_2,g_3g_4)} = F_{j-1}(g_1,g_2,1) \,.
\fe
Using this equation for general $g_1,g_2,g_3,g_4$ and also for $g_4=1$, we find:
\ie
    \delta_3 \omega_j = 1 \qquad \text{and} \qquad \delta_2 \alpha_j = \frac{\omega_j}{\omega_{j-1}} \,. \label{cocycle.condition}
\fe
Therefore, $w_j$ is a 3-cocycle and defines the anomaly index $[\omega_j]\in H^3(G,U(1))$. Moreover, this anomaly index is $j$-independent since $w_j = (\delta_2 \alpha_j) w_{j-1}$. Thus, $[\omega] \equiv [\omega_j]$ characterizes the anomaly.

The cochains have phase ambiguity due to the phase ambiguity of the fusion operators. Specifically, the phase redefinition
\ie
    \lambda_j(g_1,g_2) \mapsto \gamma_j(g_1,g_2) \lambda_j(g_1,g_2)
\fe
for $\gamma_j(g_1,g_2) \in U(1)$, leads to identifications
\ie
    \omega_j \sim \left( \delta_2 \gamma_j^{(2)} \right) \omega_j  \qquad \text{and} \qquad
    \alpha_j \sim \alpha_j \left( \delta_1 \gamma_j^{(1)} \right) \frac{\gamma_j^{(2)}}{\gamma_{j-1}^{(2)}} \,, \label{phase.redef}
\fe
where $\gamma_j^{(1)} \in C^1(G,U(1))$ and $\gamma_j^{(2)} \in C^2(G,U(1))$ are defined as
\ie
    \gamma_j^{(1)}(g_1) = \gamma_j(g_1,1) \,, \qquad \gamma_j^{(2)}(g_1,g_2) = \frac{\gamma_j(g_1,g_2)}{\gamma_j(g_1,1)} \,.
\fe

Note that the 2-cochain $\alpha_j$ can always be trivialized using the phase redefinition $\gamma_j$ satisfying
\ie
    \frac{\gamma_{j-1}}{\gamma_{j}} = \alpha_j \,. \label{trivial.alpha}
\fe
Here, we have used the fact that $\alpha_j(g_1,1)=1$. Secondly, the phase redefinition \eqref{phase.redef} only changes $\omega_j$ by an exact term and, therefore, does not change the cohomology class $[\omega_j]$. 

Finally, let us show that $F_j$ can be made trivial when the symmetry is anomaly-free, i.e., $[\omega_j] = 1$. First, notice that trivializing $\alpha_j$ would make $\omega_j$ $j$-independent because of \eqref{cocycle.condition}. Thus, after trivializing $\alpha_j$ by the phase redefinition \eqref{trivial.alpha}, we find $\omega_j = \delta_2 \beta$ for some $\beta \in C^2(G,U(1))$. Choosing the phase redefinition
\ie
    \gamma_j(g_1,g_2) = \frac{\beta(g_1,g_2)}{\beta(g_1,1)}\,,
\fe
completely trivializes all the $F$-symbols, resulting in $F_j(g_1,g_2,g_3)=1$.

As explained in \cite{Seifnashri:2023dpa}, imposing a lattice translation symmetry $T$ such that $\lambda_{j+1}(g_1,g_2) = T \lambda_{j}(g_1,g_2) T^{-1}$, we further find the $j$-independent index $[\alpha_j] \in H^2(G,U(1))$ that characterizes the mixed anomaly between $G$ and lattice translation.

\section{Topological defects and gauging \label{app:defect.gauging}}

In this Appendix, we review the gauging of anomaly-free finite-group symmetries using topological defects \cite{Seifnashri:2023dpa}.

\subsection{Defect Hamiltonians and fusion operators \label{sec:defects}}

We start by noting that the fusion operators in \eqref{fusion.op} naturally act on defect Hamiltonians. Given a symmetric Hamiltonian $H$, we construct the defect Hamiltonian
\ie
    H_g^{(j,j+1)} = U^g_{\leq j} \, H \, (U^g_{\leq j})^\dagger \,,
\fe
which corresponds to inserting a $g$ defect on link $(j,j+1)$, hence imposing a $g$-twisted boundary condition on that link. Crucially, these defects are \emph{topological} \cite{Cheng:2022sgb} in the sense that they can be moved with the unitary local operator $U_j^g = \lambda_j(g,1)$:
\ie
    H_{g}^{(j,j+1)} = U_j^g \, H_{g}^{(j-1,j)} (U_j^g)^\dagger \,.
\fe

The local terms in the defect Hamiltonian $H_g^{(j,j+1)}$ are identical to those of $H$ far away from the interval $[j{-}1,j{+}1]$. Since the defects are localized, it is unambiguous to insert multiple defects. In particular, the defect Hamiltonian in the presence of a $g$ and an $h$ defect on adjacent links is as follows:
\ie
    H_{g;h}^{(j-1,j);(j,j+1)} = U^g_{\leq j-1} U^h_{\leq j} H (U^g_{\leq j-1} U^h_{\leq j})^\dagger.
\fe
The fusion operators satisfy the defining relation
\ie
    H_{gh}^{(j,j+1)} = \lambda_j(g,h) \, H_{g;h}^{(j-1,j);(j,j+1)} \lambda_j^\dagger(g,h) \,, \label{intro.fusion.equation}
\fe
which is interpreted as the fusion of a $g$ defect on link $(j{-}1,j)$ and an $h$ defect on link $(j,j{+}1)$ into a $gh$ defect on link $(j,j{+}1)$.

\subsection{Gauging\label{app:gauging}}

To gauge the symmetry, we first enlarge the Hilbert space $\mathcal{H}$ to $\mathcal{H}\otimes \mathcal{H}_\text{links}$, where $\mathcal{H}_\text{links}=\bigotimes_{j\in\mathbb{Z}}\mathbb{C}^{|G|}$ corresponds to $|G|$-dimensional ancilla qudits on \emph{links} that are interpreted as dynamical gauge fields. We then couple these gauge fields to the Hamiltonian $H$ and obtain the Hamiltonian of the gauged theory:
\ie
    & H_\text{gauged} = \sum_{\{g_j\}} H_{\{g_j\}} \otimes \ket{\{g_j\}}\bra{\{g_j\}}_\text{links} \,, \\
    &\ket{\{g_j\}}_\text{links} = \bigotimes_{j\in \bZ} \ket{g_j}_{j+\frac12} \,,
\fe
where
\ie
{} H_{\{g_j\}} = (\cdots U^{g_{j-1}}_{\leq j-1} U^{g_j}_{\leq j} U^{g_{j+1}}_{\leq j+1} \cdots) H (\cdots U^{g_{j-1}}_{\leq j-1} U^{g_j}_{\leq j} U^{g_{j+1}}_{\leq j+1} \cdots)^\dagger
\fe
is the defect Hamiltonian, containing a $g_j$ defect on the link $(j,j{+}1)$ for each $j$. The extended Hamiltonian commutes with Gauss's law operators
\ie
    \mathsf{G}_j^g  = \sum_{a,b \in G} \raisebox{-30pt}{ \begin{tikzpicture}[scale=0.7]
\foreach \y in {1,2,3} {
  \draw (0.7,\y) -- (3.3,\y);
  % Draw lattice sites
  \foreach \x in {1,2,3} {
    \filldraw[fill=black, draw=black] (\x,\y) circle (0.06);
  }}
  % Draw defect
  \draw[color=purple, thick] (1.5,1) -- (1.5,1.5) -- (2.5,1.5);
  \draw[color=purple, thick] (2.5,1) -- (2.5,3);
  \draw[color=purple, thick] (1.5,3) -- (1.5,2.5) -- (2.5,2.5);
  % Labels
  \node [below] at (2,1) {\scriptsize $j$};
  \node [below] at (1.5,1) {\small \color{purple} $a$};
  \node [below] at (2.5,1) {\small \color{purple} $b$};
  \node [above] at (1.5,3) {\small \color{purple} $a\bar{g}$};
  \node [above] at (2.5,3) {\small \color{purple} $gb$};
\end{tikzpicture}} \, \otimes \ket{a\bar{g},gb} \bra{a,b}_{j-\frac12 , j+\frac12} =
    \sum_{a,b \in G} \lambda_j^\dagger(a\bar{g},gb) \lambda_j(a,b) \otimes \ket{a\bar{g},gb} \bra{a,b}_{j-\frac12 , j+\frac12} \,. \label{app:gauss.operator}
\fe

Note that the Gauss law operator $\mathsf{G}_j^g$ above is distinct from $\mathcal{G}_j^g$, which is defined in \eqref{gauss.operator}. As shown in \cite{Seifnashri:2023dpa} and reviewed in Appendix \ref{app:gauss.law.commute}, Gauss's law operators can be taken to commute with each other at different sites if and only if the symmetry is anomaly-free. In particular, the $F$-move identity \eqref{eq:Fmove} implies that Gauss's law operators commute at different sites, i.e., $\mathsf{G}_{j_1}^{g_1}$ commutes with $\mathsf{G}_{j_2}^{g_2}$ for $j_1 \neq j_2$.

Finally, the gauged theory is described by the extended Hamiltonian $H_\text{gauged}$ subject to the Gauss constraints
\ie
    P_j \equiv \frac{1}{|G|}\sum_{g \in G} \mathsf{G}_j^g = 1 \,. \label{gauss.constraint}
\fe
For anomaly-free symmetries, Gauss operators $P_j$ at different sites commute with each other, ensuring a consistent gauged theory.

The gauging procedure can be phrased independently of a Hamiltonian by defining the \emph{gauging map}:
\ie
    \mathsf{G} = \prod_j\left( \frac{1}{|G|} \sum_g \mathsf{G}_j^g \right) \ket{\{1\}}_\mathrm{link} 
    ~:~ \mathcal{H} \to \mathcal{H}\otimes \mathcal{H}_\mathrm{link} \,,
\fe
where $\ket{\{1\}}_\mathrm{link} = \bigotimes_j \ket{1}_{j+\frac12} \in \mathcal{H}_\mathrm{link}$. Note that the image of this map respects Gauss's law constraints. Therefore, the Hilbert space of the gauged theory $\mathcal{H}_\text{gauged}$ is identified with the subspace $\mathcal{H}\otimes \mathcal{H}_\mathrm{link}|_{P_j = 1}$. The operator algebra of the gauged theory $\mathscr{A}_\text{gauged}$ is the subalgebra of $\mathscr{A} \otimes \mathscr{A}_\mathrm{link}$ that commutes with Gauss's constraints and is subject to the equivalence relation $\mathcal{O}_1 \sim \mathcal{O}_2 ~ \Leftrightarrow ~ \forall_j: \mathcal{O}_1 P_j = \mathcal{O}_2 P_j$. Given a symmetric Hamiltonian $H$, the gauging map determines the gauged theory Hamiltonian $H_\text{gauged}$ via the relations
\ie
    H_\text{gauged} \mathsf{G} = \mathsf{G}H \,, \quad H_\text{gauged} P_j = P_j H_\text{gauged} \,.
\fe

\subsection{$[\mathsf{G}_{j_1}^{g_1},\mathsf{G}_{j_2}^{g_2}]=0$ for $j_1 \neq j_2$ \label{app:gauss.law.commute}}

Here, we review the fact that for anomaly-free symmetries, Gauss's law operators commute at different sites \cite{Seifnashri:2023dpa}. Specifically, we show $[\mathsf{G}_{j_1}^{g_1},\mathsf{G}_{j_2}^{g_2}]=0$ and $[\mathcal{G}_{j_1}^{g_1},\mathcal{G}_{j_2}^{g_2}]=0$ for $j_1 \neq j_2$.

Note that for $|j-j'| \geq 2$, $\mathcal{G}_{j}^g$ commutes with $\mathcal{G}_{j'}^h$ since $\lambda_j(\cdot\hspace{1pt},\cdot)$ commutes with $\lambda_{j'}(\cdot\hspace{1pt},\cdot)$. Thus, the only non-trivial relation to show is $\mathcal{G}_{j-1}^g \mathcal{G}_j^h = \mathcal{G}_j^h \mathcal{G}_{j-1}^g$ and $\mathsf{G}_{j-1}^g \mathsf{G}_j^h = \mathsf{G}_j^h \mathsf{G}_{j-1}^g$. By taking various matrix elements, they are equivalent to showing:
\ie
	\raisebox{-50pt}{
\begin{tikzpicture}[scale=0.7]
 \foreach \y in {0,1,2,3,4}{
 \draw (0.7,\y) -- (4.3,\y);
  % Draw lattice sites
  \foreach \x in {1,2,3,4} {
    \filldraw[fill=black, draw=black] (\x,\y) circle (0.045);
  }}
  
  % Draw defect
  \draw[color=purple, thick] (1.5,0) -- (1.5,0.5) -- (2.5,0.5);
  \draw[color=purple, thick] (1.5,4) -- (1.5,1.5) -- (2.5,1.5);
  \draw[color=purple, thick] (2.5,0) -- (2.5,2.5);
  \draw[color=purple, thick] (2.5,2.5) -- (3.5,2.5);
  \draw[color=purple, thick] (3.5,0) -- (3.5,4);
  \draw[color=purple, thick] (3.5,3.5) -- (2.5,3.5) -- (2.5,4);

  % Labels
  \node [below] at (3,0) {\scriptsize$j$};
  \node [below] at (1.5,0) {\small \color{purple} $g_1$};
  \node [below] at (2.5,0) {\small\color{purple} $g_2$};
  \node [below] at (3.5,0) {\small\color{purple} $g_3$};
  \node [above] at (1.5,4) {\small \color{purple} $g'_1$};
  \node [above] at (2.5,4) {\small\color{purple} $g'_2$};
  \node [above] at (3.5,4) {\small\color{purple} $g'_3$};
\end{tikzpicture}}\, \stackrel{?}{=}
\raisebox{-50pt}{
\begin{tikzpicture}[scale=0.7]
 \foreach \y in {0,1,2,3,4}{
 \draw (-0.3,\y) -- (3.3,\y);
  % Draw lattice sites
  \foreach \x in {0,1,2,3} {
    \filldraw[fill=black, draw=black] (\x,\y) circle (0.045);
  }}
  
  % Draw defect
  \draw[color=purple, thick] (1.5,0) -- (1.5,0.5) -- (2.5,0.5);
  \draw[color=purple, thick] (1.5,4) -- (1.5,1.5) -- (2.5,1.5);
  \draw[color=purple, thick] (2.5,0) -- (2.5,4);
  \draw[color=purple, thick] (0.5,0) -- (0.5,2.5) -- (1.5,2.5);
\draw[color=purple, thick] (1.5,3.5) -- (0.5,3.5) -- (0.5,4);
  
  % Labels
  \node [below] at (2,0) {\scriptsize$j$};
  \node [below] at (0.5,0) {\small \color{purple} $g_1$};
  \node [below] at (1.5,0) {\small \color{purple} $g_2$};
  \node [below] at (2.5,0) {\small\color{purple} $g_3$};
  \node [above] at (0.5,4) {\small \color{purple} $g_1'$};
  \node [above] at (1.5,4) {\small \color{purple} $g_2'$};
  \node [above] at (2.5,4) {\small\color{purple} $g_3'$};
\end{tikzpicture}} \,, \label{app:gauss.commute}
\fe
for all $g_1,g_2,g_3\in G$ where $g_1' = g_1\bar{g}$, $g_2'=g g_2 \bar{h}$, and $g_3'= h g_3$. Below, we will show this relation by applying a sequence of $F$-moves and using \eqref{eq:Fmove}.

By multiplying both sides of relation \eqref{app:gauss.commute} with $\lambda_j(g_1',g_2'g_3')\lambda_{j-1}(g_1',1)\lambda_j(g_2',g_3')$, it is simplified to
\ie
\raisebox{-70pt}{
\begin{tikzpicture}[scale=0.7]
 \foreach \y in {0,1,2,3,4,5}{
 \draw (0.7,\y) -- (4.3,\y);
  % Draw lattice sites
  \foreach \x in {1,2,3,4} {
    \filldraw[fill=black, draw=black] (\x,\y) circle (0.05);
  }}
  
  % Draw defect
  \draw[color=purple, thick] (1.5,0) -- (1.5,0.5) -- (2.5,0.5);
  \draw[color=purple, thick] (3.5,4.5) --(2.5,4.5) --(2.5,3.5) --(1.5,3.5) -- (1.5,1.5) -- (2.5,1.5);
  \draw[color=purple, thick] (2.5,0) -- (2.5,2.5);
  \draw[color=purple, thick] (2.5,2.5) -- (3.5,2.5);
  \draw[color=purple, thick] (3.5,0) -- (3.5,5);
  
  % Labels
  \node [left] at (1.5,2.5) {\footnotesize \color{purple} $g_1'$};
  \node [below] at (1.5,0) {\footnotesize \color{purple} $g_1$};
  \node [below] at (2.5,0) {\footnotesize\color{purple} $g_2$};
  \node [below] at (3.5,0) {\footnotesize\color{purple} $g_3$};
  \node [above] at (3.5,5) {\footnotesize \color{purple} $g'_1g'_2g'_3$};
\end{tikzpicture}} \,\stackrel{?}{=}
\raisebox{-70pt}{
\begin{tikzpicture}[scale=0.7]
 \foreach \y in {0,1,2,3,4,5,6,7}{
 \draw (-0.3,\y) -- (3.3,\y);
  % Draw lattice sites
  \foreach \x in {0,1,2,3} {
    \filldraw[fill=black, draw=black] (\x,\y) circle (0.05);
  }}
  
  % Draw defect
  \draw[color=purple, thick] (1.5,0) -- (1.5,0.5) -- (2.5,0.5);
  \draw[color=purple, thick] (2.5,4.5) -- (1.5,4.5) -- (1.5,1.5) -- (2.5,1.5);
  \draw[color=purple, thick] (2.5,0) -- (2.5,7);
  \draw[color=purple, thick] (0.5,0) -- (0.5,2.5) -- (1.5,2.5);
  \draw[color=purple, thick] (1.5,3.5) -- (0.5,3.5) -- (0.5,4.5) --(0.5,5.5) --(1.5,5.5) --(1.5,6.5) --(2.5,6.5) ;
  
  % Labels
  \node [above left] at (0.65,3.85) {\footnotesize \color{purple} $g_1'$};
  \node [above left] at (1.65,3.85) {\footnotesize \color{purple} $g_2'$};
  \node [above right] at (2.5,3.85) {\footnotesize \color{purple} $g_3'$};
  \node [below] at (0.5,0) {\footnotesize \color{purple} $g_1$};
  \node [below] at (1.5,0) {\footnotesize \color{purple} $g_2$};
  \node [below] at (2.5,0) {\footnotesize\color{purple} $g_3$};
  \node [above] at (2.5,7) {\footnotesize\color{purple} $g'_1g'_2g'_3$}; 
\end{tikzpicture}} \,.
\fe
By using $F$-move identities \eqref{eq:Fmove} on the top, it becomes
\ie
\raisebox{-30pt}{
\begin{tikzpicture}[scale=0.7]
 \foreach \y in {0,1,2}{
 \draw (0.7,\y) -- (4.3,\y);
  % Draw lattice sites
  \foreach \x in {1,2,3,4} {
    \filldraw[fill=black, draw=black] (\x,\y) circle (0.045);
  }}
  
  % Draw defect
	\draw[color=purple, thick] (3.5,0) -- (3.5,2);
	\draw[color=purple, thick] (2.5,0) -- (2.5,1.5)-- (3.5,1.5);
	\draw[color=purple, thick] (1.5,0) -- (1.5,0.5)-- (2.5,0.5);

  % Labels
  \node [below] at (3,0) {\scriptsize$j$};
  \node [below] at (1.5,0) {\footnotesize \color{purple} $g_1$};
  \node [below] at (2.5,0) {\footnotesize\color{purple} $g_2$};
  \node [below] at (3.5,0) {\footnotesize\color{purple} $g_3$};
  \node [above] at (3.5,2) {\footnotesize \color{purple} $g'_1g'_2g'_3$};
\end{tikzpicture}} \,\stackrel{?}{=} 
\raisebox{-50pt}{
\begin{tikzpicture}[scale=0.7]
 \foreach \y in {0,1,2,3,4}{
 \draw (-0.3,\y) -- (3.3,\y);
  % Draw lattice sites
  \foreach \x in {0,1,2,3} {
    \filldraw[fill=black, draw=black] (\x,\y) circle (0.045);
  }}
  
  % Draw defect
  \draw[color=purple, thick] (1.5,0) -- (1.5,0.5) -- (2.5,0.5);
  \draw[color=purple, thick] (2.5,3.5) -- (1.5,3.5) -- (1.5,1.5) -- (2.5,1.5);
  \draw[color=purple, thick] (2.5,0) -- (2.5,4);
  \draw[color=purple, thick] (0.5,0) -- (0.5,2.5) -- (1.5,2.5);
  
  % Labels
  \node [above right] at (2.5,2) {\footnotesize \color{purple} $g_3'$};
  \node [below] at (2,0) {\scriptsize$j$};
  \node [below] at (0.5,0) {\footnotesize \color{purple} $g_1$};
  \node [below] at (1.5,0) {\footnotesize \color{purple} $g_2$};
  \node [below] at (2.5,0) {\footnotesize\color{purple} $g_3$};
  \node [above] at (2.5,4) {\footnotesize\color{purple} $g'_1g'_2g'_3$};
\end{tikzpicture}} \,,
\fe
which itself follows from applying two additional $F$-move identities.

\section{Relation between gauging and disentangling \label{app:rel.gauging}}

In this appendix, we provide a physical interpretation of the disentangler $\mathcal{W}$. Specifically, we show that the disentangler implements an isomorphism between a `theory' $\mathcal{T}$ with an anomaly-free $G$-symmetry and the theory $\mathcal{T}' = \frac{\mathcal{T} \times G\text{-SSB}}{G}$, which is obtained by stacking $\mathcal{T}$ with a $G$-SSB phase and subsequently gauging the diagonal $G$-symmetry. Throughout, we use the term \emph{theory} to refer to a Hamiltonian system.

\subsection{Gauging $U^g \otimes \prod_j R_j^{\bar{g}}$}

To gauge a symmetry, as reviewed in Appendix \ref{app:gauging}, we couple the Hamiltonian to gauge fields and impose Gauss constraints $\frac{1}{|G|}\sum_g \mathsf{G}_j^g=1$, where $\mathsf{G}_j^g$ is the Gauss law operator at site $j$ given in \eqref{app:gauss.operator}.

Here, we modify Gauss's law operators from $\mathsf{G}_j^g$ to 
\ie
    \mathcal{G}_j^g = \sum_{a,b\in G}  \left(  \lambda_j^\dagger(a\bar{g},gb)\lambda_j(a,b) \otimes \sum_{s\in G} \ket{s\bar{a},s\bar{g},sb} \bra{s\bar{a},s,sb}_{j-1,j,j+1} \right) \,. \label{modified.gauss}
\fe
As we will show below, the modified Gauss law corresponds to stacking the system with a $G$-SSB and gauging the diagonal symmetry generated by $U^g \otimes \prod_j R_j^{\bar{g}}$ instead of gauging the original symmetry $U^g$. Here, $R_j^{g}=\sum_s \ket{sg}\bra{s}_j$ is the right multiplication operator at site $j$. 

\subsubsection*{Stacking with an SSB state}

We begin with theory $\mathcal{T}$ that corresponds to a Hamiltonian $H$ that has an anomaly-free $G$-symmetry generated by symmetry operators $U^g$.
Stacking with a $G$-SSB phase amounts to extending the Hilbert space from $\mathcal{H}$ to $\mathcal{H} \otimes \mathcal{H}'$, and extending the Hamiltonian $H$ to
\ie
     H \otimes \mathbbm{1} - \lambda \mathbbm{1} \otimes \left( \sum_{j,g} \ket{g,g}\bra{g,g}_{j,j+1}\right) \,.
\fe

We take $\lambda = +\infty$, which is equivalent to working with the Hamiltonian $H \otimes \mathbbm{1}$ subject to commuting local constraints $P_{j+\frac12} =1$, where
\ie
    P_{j+\frac12} = \sum_{s \in G} \ket{s,s}\bra{s,s}_{j,j+1} \,.
\fe
We will refer to this theory as $\mathcal{T} \times G\text{-SSB}$.

\subsubsection*{Gauging the diagonal symmetry}
Gauging the diagonal $G$-symmetry $U^g \otimes \prod_j R_j^{\bar{g}}$ is implemented by the Gauss law operators:
\ie
    \widetilde{P}_j \equiv \frac{1}{|G|}\sum_g \mathsf{G}_j^g \otimes R_j^{\bar{g}} = 1 \,,
\fe
and the gauging map
\ie
    \widetilde{\mathsf{G}} = \prod_j \widetilde{P}_j \ket{\{1\}}_\text{link} \,.
\fe
Specifically, the gauged theory is described by the Hamiltonian $\widetilde{H}$ satisfying
\ie
    \widetilde{H} \, \widetilde{\mathsf{G}} = \widetilde{\mathsf{G}} \, (H \otimes \mathbbm{1})\,, \qquad [\widetilde{H},\widetilde{P}_j]=[\widetilde{H},\widetilde{P}_{j+\frac12}]=0 \,,
\fe
where
\ie
    \widetilde{P}_{j+\frac12} = \sum_{g,s \in G} \ket{s\bar{g},g,s}\bra{s\bar{g},g,s}_{j,j+\frac12,j+1}
\fe
is the image of the constraint $P_{j+\frac12}$ under the gauging map, i.e., $\widetilde{P}_{j+\frac12} \widetilde{\mathsf{G}} = \widetilde{\mathsf{G}} P_{j+\frac12}$. Note that $\widetilde{P}_{j+\frac12}$ commutes with Gauss's constraints $\widetilde{P}_{j'}$.
 
\subsubsection*{Gauge fixing}

By performing the local unitary transformation
\ie
    \mathsf{U} = \prod_j \left( \sum_{g\in G} R_{j-\frac12}^{\bar{g}} L_{j+\frac12}^{g} \,{\ket{g}}{\bra{g}}_j \right)\,,
\fe
we can trivialize the local constraints $\widetilde{P}_{j+\frac12} = 1$ to
\ie
    \mathsf{U} \widetilde{P}_{j+\frac12} \mathsf{U}^\dagger = \ket{1}\bra{1}_{j+\frac12} = 1 \,.
\fe
Therefore, applying the unitary transformation $\mathsf{U}$ decouples the degrees of freedom on the links and ``gauge fixes'' them to $\ket{\{1\}}_\text{link}$. 

Moreover, by defining
\ie
    H' &= \bra{\{1\}} \mathsf{U} \widetilde{H} \mathsf{U}^\dagger \ket{\{1\}}_\text{link} \,, \\
    P'_j &= \bra{\{1\}} \mathsf{U} \widetilde{P}_j \mathsf{U}^\dagger \ket{\{1\}}_\text{link} \,,\\
    \mathsf{G}' &= \bra{\{1\}} \mathsf{U} \widetilde{\mathsf{G}} \,,
\fe
the gauged theory Hamiltonian is given by $H'$ satisfying
\ie
    H' \mathsf{G}' = \mathsf{G}' (H \otimes \mathbbm{1}) \,, \qquad [H',P'_j]=0 \,.
\fe
The modified Gauss constraints are
\ie
    P'_j = \frac{1}{|G|} \sum_{g \in G} \mathcal{G}^g_j\,.
\fe
This establishes that the (modified) Gauss law operators $\mathcal{G}^g_j$ of \eqref{modified.gauss} indeed implement gauging of the diagonal $G$-symmetry of the theory $\mathcal{T}$ stacked with a $G$-SSB state. We will refer to this final theory as $\mathcal{T}' = \frac{\mathcal{T} \times G\text{-SSB}}{G}$.

\subsection{Physical interpretation of the disentangler \label{sec:interpretation}}

We interpret the disentangler $\mathcal{W}$ as the unitary transformation that implements the duality/isomorphism
\ie
    \mathcal{T} \simeq \frac{\mathcal{T} \times G\textrm{-SSB}}{G} \,.
\fe
Intuitively, gauging identifies the original symmetry $\{U^g\}$ with the on-site symmetry $\{\mathcal{X}^g\}$, therefore making it on-site.

More specifically, the theory $\mathcal{T}$ is unitarily related to the gauged theory $\frac{\mathcal{T} \times G\textrm{-SSB}}{G}$ via the disentangler $\mathcal{W}$ in the sense that:
\ie
    \mathcal{W}^\dagger \big( \mathbbm{1} \otimes R^{\bar{g}}_j \big) \mathcal{W} &= \mathcal{G}^g_j \,, \\
    \mathcal{W}^\dagger \big( H \otimes \mathbbm{1} \big) \mathcal{W} &= H' \,, \\
    \mathcal{W}^\dagger \big( U^g \otimes \mathcal{X}^g \big) \mathcal{W} &= \mathbbm{1} \otimes \mathcal{X}^g \,.
\fe
Note that the original theory $\mathcal{T}$ with Hamiltonian $H$ and symmetry $U^g$ is isomorphic to the system with Hamiltonian $H\otimes \mathbbm{1}$ subject to the local constraints $\frac{1}{|G|} \sum_g R^{\bar{g}}_j=1$ and with symmetry $U^g \otimes \mathcal{X}^g$. The constraint $R^{\bar{g}}_j=1$ freezes the ancilla system into a product state and recovers the original theory.

One benefit of the interpretation above is that it suggests a generalization to higher dimensions. In particular, the relation
\ie
    \mathcal{W}^\dagger \big( \mathbbm{1} \otimes R^{\bar{g}}_j \big) \mathcal{W} &= \mathcal{G}^g_j \,,
\fe
and \eqref{modified.gauss} naturally leads to an expression for the disentangler $\mathcal{W}$.

\bibliography{ref}

\end{document}